\documentclass[prl,twocolumn,showpacs]{revtex4}
\usepackage{graphicx}
\usepackage{graphics}
\usepackage{amsfonts}
\usepackage{amsmath}
\usepackage{epsfig}

\input{tcilatex}

\begin{document}

\title{Anomalous particle-number fluctuations in a three-dimensional
interacting Bose-Einstein condensate}
\author{Shujuan Liu$^{1,2}$, Hongwei Xiong$^{1,2}$, Guoxiang Huang$^{3}$,
and Zhijun Xu$^{2}$}


\address{$^{1}$Wuhan Institute of Physics and Mathematics, The Chinese Academy of
Sciences, Wuhan 430074, People¡¯s Republic of China} 
\address{$^{2}$Department of Applied Physics, Zhejiang University of
Technology, Hangzhou, 310032, China} 
\address{$^{3}$Department of Physics and Key Laboratory for Optical and Magnetic
         Resonance Spectroscopy, \\
         East China Normal University, Shanghai, 200062, China}


\begin{abstract}
The particle-number fluctuations originated from collective excitations are
investigated for a three-dimensional, repulsively interacting Bose-Einstein
condensate (BEC) confined in a harmonic trap. The contribution due to the
quantum depletion of the condensate is calculated and the explicit
expression of the coefficient in the formulas denoting the particle-number
fluctuations is given. The results show that the particle-number
fluctuations of the condensate follow the law $<\delta N_0^2> \sim N^{22/15}$
and the fluctuations vanish when temperature approaches to the BEC critical
temperature. \vspace{0.5cm}
\end{abstract}

\maketitle


The remarkable experimental realization of Bose-Einstein condensation in
trapped, weakly interacting atomic gases has stimulated intensive
theoretical and experimental researches on Bose condensed gases\cite{PET}.
In particular, much attention has been paid to the study of the
particle-number fluctuations of Bose-Einstein condensates (BECs)\cite{CHE}.
This is mainly due to the fact that the particle-number fluctuations play a
central role in the understanding of the statistical properties of BEC. In
addition, the particle-number fluctuations of a condensate may result in the
fluctuations of chemical potential and thus lead to the phase diffusion of
condensate\cite{PHASE}. Therefore, the study on the particle-number
fluctuations of BEC is not only of an intrinsic theoretical interest but
also useful to understand and control the coherent property of BEC. On the
other hand, a clear theoretical understanding will be helpful to guide a
direct experimental observation on the particle-number fluctuations in a BEC.

Up to now there exist a lot of theoretical works exploring the property of
the particle-number fluctuations in BECs. In an ideal Bose gas the
particle-number fluctuations $<\delta N_0^2>$ $\equiv <N_0^2>-<N_0>^2$ have
been studied rather thoroughly in a homogeneous case (e. g. in a box)\cite%
{BOX} and an inhomogeneous case (e. g. in a trap)\cite{TRAP}. The role of
interatomic interaction in the particle-number fluctuations is an important
theoretical problem and hence intensive theoretical investigations have been
given \cite{CHE,GIO,KOC,MEI,IDZ,ILL,JAK,BHA,XIONG1,XIONG2,XIONG3,XIONG2D}.
The behavior of the particle-number fluctuations is described by the value
of the power $\gamma$ in the expression $<\delta N_0^2>\sim N^{\gamma}$ with 
$N$ being the total number of particles. Refs.\cite{GIO,KOC,XIONG1} give the
result of \textit{anomalous} fluctuations with $\gamma=4/3$, while Ref.\cite%
{IDZ} argue that the fluctuations should be normal with $\gamma=1$.

For the temperature far below the critical temperature of BEC, the
collective excitations created from condensate play an important role in the
particle-number fluctuations. The physical reason is that due to the
interatomic interaction the creation of the collective excitations induces a
change of the particle number in the condensate. In the last few years, the
particle-number fluctuations originated from collective excitations have
been studied for three-dimensional (3D)\cite{GIO,KOC,MEI,XIONG1,XIONG2} and
2D\cite{XIONG3} weakly interacting Bose-condensed gases. In a pioneering
work by Giorgini et al \cite{GIO}, the particle-number fluctuations are
investigated within a traditional particle-number-nonconserving Bogoliubov
method. Kocharovsky et al\cite{KOC} extended the results of \cite{GIO} by
using a particle-number-conserving operator formalism. The scale behavior of
the interacting condensate in a box was investigated in Ref.\cite{MEI} with
an arbitrary atomic interaction.

For a Bose-condensed gas confined in a magnetic trap, the total number of
particles $N$ of the system is conserved and hence a canonical ensemble is
appropriate to investigate the particle-number fluctuations in the
condensate. Directly from the canonical partition function of the system and
by using a developed saddle-point method, a systematic approach was proposed
by the present authors\cite{XIONG1,XIONG2,XIONG3,XIONG2D} for investigating
the role of collective excitations on the particle-number fluctuations.
Within the canonical ensemble a general method has been given recently for
studying the thermodynamic properties of Bose-condensed gases based on the
calculation of the probability distribution function\cite{XIONG1,XIONG2}. In
Ref.\cite{XIONG3}, the theory in \cite{XIONG1,XIONG2} is developed to
calculate the particle-number fluctuations due to collective excitations by
including the effect of quantum depletion.

For a 3D Bose-condensed gas confined in a magnetic trap, although the effect
of the quantum depletion of condensate is discussed by Giorgini \textit{et
al }\cite{GIO} within the particle-number-nonconserving Bogoliubov approach,
explicit expression in the formula denoting the particle-number fluctuations
by the quantum depletion is not provided. In the present work, we shall
carefully calculate the particle-number fluctuations due to collective
excitations within a canonical ensemble. We take into account the effect of
quantum depletion and finite-size of the condensate.

We first give a brief description for the method developed in Ref. \cite%
{XIONG3} . For a condensate confined in a 3D magnetic trap, the collective
excitations generated from the condensate can be described by three quantum
numbers $n$ (the number of radial modes), $l$ (the magnitude of the total
angular momentum), and $m$ (the $z$-component of the angular momentum).
Assuming that $N_{nlm}^{B}$ is the number of the collective excitations
indexed by the quantum numbers $nlm$ and $N_{0}$ is the number of particles
in the condensate, the canonical partition function of the system takes the
form 
\begin{equation}
Z\left[ N\right] =\sum^{\prime }\exp \left[ -\beta \left(
E_{0}+\sum_{nlm\neq 0}N_{nlm}^{B}\varepsilon _{nlm}\right) \right] ,
\label{par}
\end{equation}%
where the prime in the summation represents the condition that the total
number of atoms in the system should be conserved within the canonical
ensemble, $E_{0}$ is the energy of the condensate which can be regarded as a
ground-state energy of the system. Using the Bogoliubov theory \cite%
{BOG,GIOBO} and a saddle-point method developed in \cite%
{XIONG1,XIONG2,XIONG3}, for the temperature below the BEC critical
temperature the normalized probability distribution function reads\cite%
{XIONG3} 
\begin{equation}
G_{n}\left( N,N_{0}\right) =A_{n}\exp \left[ -\frac{\left(
N_{0}-N_{0}^{p}\right) ^{2}}{2\Xi }\right] ,  \label{prabability1}
\end{equation}%
where $A_{n}$ is the normalization constant and $N_{0}^{p}$ is the most
probable value of the atomic number in the condensate. The quantity $\Xi $
is given by 
\begin{eqnarray}
&&\Xi =\sum_{nlm\neq 0}\left[ \left( \int u_{nlm}^{2}dV+\int
v_{nlm}^{2}dV\right) ^{2}\left( \frac{k_{B}T}{\varepsilon _{nlm}}\right)
^{2}\right.  \notag \\
&&+2\left( \int u_{nlm}^{2}dV+\int v_{nlm}^{2}dV\right) \int
v_{nlm}^{2}dV\left( \frac{k_{B}T}{\varepsilon _{nlm}}\right)  \notag \\
&&\hspace{0.8cm}\left. +\left( \int v_{nlm}^{2}dV\right) ^{2}\right] ,
\label{xi}
\end{eqnarray}%
where $u_{nlm}$, $v_{nlm}$ and $\varepsilon _{nlm}$ (the energy of the
collective mode $nlm$) are determined by the following coupled equations:

$${
\left( -\frac{\hbar ^{2}}{2m}\bigtriangledown ^{2}+V_{ext}\left( \mathbf{r}%
\right) -\mu +2gn\left( \mathbf{r}\right) \right) u_{nlm}+gn_{0}\left( 
\mathbf{r}\right) v_{nlm}
}$$

\begin{equation}
=\varepsilon _{nlm}u_{nlm},  \label{uvenergy1}
\end{equation}

$${
\left( -\frac{\hbar ^{2}}{2m}\bigtriangledown ^{2}+V_{ext}\left( \mathbf{r}%
\right) -\mu +2gn\left( \mathbf{r}\right) \right) v_{nlm}+gn_{0}\left( 
\mathbf{r}\right) u_{nlm}
}$$

\begin{equation}
=-\varepsilon _{nlm}v_{nlm},  \label{uvenergy2}
\end{equation}%
where $V_{ext}\left( \mathbf{r}\right) $ is the external potential confining
the Bose gas, $\mu $ and $g$ are respectively the chemical potential of the
system and the coupling constant; $n\left( \mathbf{r}\right) $ and $%
n_{0}\left( \mathbf{r}\right) $ are the density distributions of the Bose
gas and the condensate, respectively.

For temperature far below the critical temperature, we have $N_{0}^{p}>>1$.
In this situation, from Eq. (\ref{prabability1}) the fluctuations of the
condensate contributed from the collective excitations are given by 
\begin{equation}
\left\langle \delta ^{2}N_{0}\right\rangle =\left\langle
N_{0}^{2}\right\rangle -\left\langle N_{0}\right\rangle ^{2}=\Xi .
\label{fluc}
\end{equation}
For Eqs. (\ref{uvenergy1}) and (5), $u_{nlm}$ and $v_{nlm}$ can be
approximated as\cite{GRIFFIN} 
\begin{equation}
u_{nlm}\approx -v_{nlm}\approx i\sqrt{\frac{gn_{0}\left( \mathbf{r}\right) }{%
2\varepsilon _{nlm}}}\chi _{nlm}.  \label{uv}
\end{equation}

For a Bose gas confined in an isotropic harmonic potential with angular
frequency $\omega $, $\chi _{nlm}$ and $\varepsilon _{nlm}(=\hbar \omega
_{nlm})$ are determined by the eigen equation 
\begin{equation}
-\frac{\omega ^{2}}{2}\nabla \cdot \left[ \left( R^{2}-r^{2}\right) \nabla
\chi _{nlm}\right] =\omega _{nlm}^{2}\chi _{nlm},  \label{xixi}
\end{equation}
where $R$ is the radius of the condensate, determined by the chemical
potential $\mu $ of the system through $\mu =m\omega ^{2}R^{2}/2$. In Eq. (%
\ref{xixi}), $\omega _{nlm}$ and $\chi _{nlm}$ are found to be\vspace{1pt} 
\cite{STR} 
\begin{equation}
\omega _{nlm}=\omega \left( 2n^{2}+2nl+3n+l\right) ^{1/2},  \label{frequency}
\end{equation}
and 
\begin{equation}
\chi _{nlm}=A_{nl}P_{l}^{\left( 2n\right) }\left( \frac{r}{R}\right)
r^{l}Y_{lm}\left( \theta ,\varphi \right) \Theta \left( R-r\right) ,
\label{xixi1}
\end{equation}
where $A_{nlm}$ is the normalization constant determined by $\int \left|
\chi _{nlm}\right| ^{2}dV=1$ and $\Theta \left( x\right) $ is a step
function. In Eq. (\ref{xixi1}), $P_{l}^{\left( 2n\right) }\left( x\right)
=\sum_{k=0}^{n}\alpha _{2k}x^{2k}$ is a polynomial with coefficients
satisfying the recurrence relation: $\alpha _{2k+2}=-\alpha _{2k}\left(
n-k\right) \left( 2l+2k+2n+3\right) /\left( k+1\right) \left( 2l+2k+3\right) 
$ with $\alpha _{0}=1$.

With the above formulas we now calculate the particle-number fluctuations in
the condensate. Substituting the above results into Eqs. (\ref{xi}) and (\ref%
{fluc}), we obtain the particle-number fluctuations due to collective
excitations: 
\begin{equation}
\left\langle \delta ^{2}N_{0}\right\rangle =\Re _{1}+\Re _{2}+\Re _{3},
\label{analyticalfluc}
\end{equation}
where 
\begin{equation}
\Re _{1}=\lambda _{1}\left( \frac{\mu }{\hbar \omega }\right) ^{2}\left( 
\frac{k_{B}T}{\hbar \omega }\right) ^{2},
\end{equation}
\begin{equation}
\Re _{2}=\lambda _{2}\left( \frac{\mu }{\hbar \omega }\right) ^{2}\frac{%
k_{B}T}{\hbar \omega },
\end{equation}
and 
\begin{equation}
\Re _{3}=\frac{\lambda _{3}}{4}\left( \frac{\mu }{\hbar \omega }\right) ^{2}.
\end{equation}
The coefficients $\lambda _{1}$, $\lambda _{2}$, $\lambda _{3}$ are given by

\begin{equation*}
{\ \lambda _{1}=\sum_{nl\neq 0}\frac{\left( 2l+1\right) \vartheta _{nl}^{2}}{%
\left( 2n^{2}+2nl+3n+l\right) ^{2}},}
\end{equation*}

\begin{equation*}
{\ \lambda _{2}=\sum_{nl\neq 0}\frac{\left( 2l+1\right) \vartheta _{nl}^{2}}{%
\left( 2n^{2}+2nl+3n+l\right) ^{3/2}},}
\end{equation*}
\begin{equation}
\lambda _{3}=\sum_{nl\neq 0}\frac{\left( 2l+1\right) \vartheta _{nl}^{2}}{%
\left( 2n^{2}+2nl+3n+l\right) },  \label{gamma}
\end{equation}
with 
\begin{equation}
\vartheta _{nl}=\frac{\int_{0}^{1}\left( 1-x^{2}\right) \left( P_{l}^{\left(
2n\right) }\left( x\right) \right) ^{2}x^{2l+2}dx}{\int_{0}^{1}\left(
P_{l}^{\left( 2n\right) }\left( x\right) \right) ^{2}x^{2l+2}dx}.
\label{beta}
\end{equation}
The factor $\left( 2l+1\right) $ in ${\lambda _{1}}$, ${\lambda _{2}}$ and ${%
\lambda _{3}}$ is due to the $\left( 2l+1\right) -$fold degeneracy of the
angular momentum. The numerical results of the above parameters are ${%
\lambda _{1}=0.47}$, ${\lambda _{2}=0.94}$, ${\lambda _{3}=4.96}$. Note that
the coefficients in the particle-number fluctuations of the condensate is
different from the result given in Refs. \cite{GIO} and \cite{XIONG1}.

In Eq. (\ref{analyticalfluc}), $\Re _{1}$ is the leading term of the
particle-number fluctuations, while $\Re _{3}$ shows the particle-number
fluctuations due to the quantum depletion of the condensate which do not
vanish even at zero temperature.

For the Bose-condensed gas in the harmonic potential, the chemical potential
is given by 
\begin{equation}
\mu =\frac{\hbar \omega }{2}\left( \frac{15N_{0}a}{a_{ho}}\right) ^{2/5},
\end{equation}
with $a$ being the $s-$wave scattering length and $a_{ho}=\sqrt{\hbar
/m\omega }$ the harmonic oscillator length. Using the expression of the
critical temperature of an ideal Bose gas, $T_{c}^{0}$=$\hbar \omega \left(
N/\zeta \left( 3\right) \right) ^{1/3}/k_{B}$, and introducing the
dimensionless parameters $t=T/T_{c}^{0}$ and $\sigma =a/a_{ho}$, we obtain
the leading term of $\Re _{1} $: 
\begin{equation}
\Re _{1}=0.91t^{2}\left( 1-t^{3}\right) ^{4/5}\sigma ^{4/5}N^{22/15}.
\label{leading2}
\end{equation}
We see that the leading term of the particle-number fluctuations in the
condensate exhibits an anomalous behavior of $\left\langle \delta
^{2}N_{0}\right\rangle \sim N^{22/15}$.

It is straightforward to obtain the results of $\Re _{2}$ and $\Re _{3}$,
which are given by 
\begin{equation}
\Re _{2}=1.93t\left( 1-t^{3}\right) ^{4/5}\sigma ^{4/5}N^{17/15},
\end{equation}
and 
\begin{equation}
\Re _{3}=2.70\left( 1-t^{3}\right) ^{4/5}\sigma ^{4/5}N^{4/5}.
\end{equation}

We now make some remarks on the results obtained above: (i)The anomalous
behavior of the particle-number fluctuations of the condensate predicted by (%
\ref{leading2}) is obtained when the particle number of the system is
finite. In the thermodynamics limit of the Bose gas in the harmonic
potential, i. e. letting $N\rightarrow \infty $ and $\omega \rightarrow 0$
while keeping $N\omega ^{3}$ constant \cite{RMP}, the anomalous behavior of
the form $\left\langle \delta ^{2}N_{0}\right\rangle \sim N^{4/3}$ given in
Refs.\cite{GIO,XIONG1} can be obtained. (ii)Different from the result
obtained in Ref.\cite{GIO} where only the low-temperature behavior $%
T<<T_{c}^{0}$ is taken into account, in the present work the temperature
dependence of the fluctuations is valid for the temperature region below the
critical temperature. When the temperature $T$ approaches the BEC critical
temperature $T_{c}^{0}$, the particle number in the condensate approaches
zero and hence the fluctuations originated from the collective excitations
vanish, as shown in Eq. (\ref{leading2}). (iii)For a higher temperature the
collective excitations play only a second role and one must consider the
contribution from single-particle excitations. The particle-number
fluctuations contributed by the single-particle excitations show the
behavior of $\left\langle \delta ^{2}N_{0}\right\rangle \sim N$ \cite{XIONG1}%
. (iv)From Eq. (\ref{leading2}), we see that the confining potential has the
effect of increasing the particle number fluctuations. It is easy to know
from Eq. (\ref{leading2}) that $\Re _{1}\sim \omega ^{2/5}$.

In conclusion, we have investigated the particle-number fluctuations
originated from collective excitations in a three-dimensional, repulsively
interacting Bose-Einstein condensate (BEC) confined in a harmonic potential.
We have carefully calculated the contribution to the fluctuations due to the
quantum depletion and provided the explicit expression of the coefficient in
the formulas denoting the particle-number fluctuations. Our results show
that the particle-number fluctuations of the condensate due to the
collective excitations display an anomalous behavior of the form $<\delta
N_0^2> \sim N^{22/15}$. When $T\rightarrow T_c^0$ such fluctuations approach
to zero and hence the fluctuations due to single-particle excitations become
dominant. It is possible that the anomalous behavior predicted in this work
could be observed experimentally by means of, e. g., the scattering of short
and nonresonant laser pulses on a BEC \cite{LEWENSTEIN}.

This work was supported by NSFC under Grants Nos. 10205011 and 10274021.


\end{document}